%% file: main.tex
\documentclass[screen,acmsmall,nonacm]{acmart}

\usepackage{cleveref}
\usepackage{xspace}
\usepackage{fancyvrb}

\definecolor{gray50}{gray}{.5}
\definecolor{gray40}{gray}{.6}
\definecolor{gray30}{gray}{.7}
\definecolor{gray20}{gray}{.8}
\definecolor{gray10}{gray}{.9}
\definecolor{gray05}{gray}{.95}

\newcommand{\ie}{i.e.,\xspace}

\newlength\Linewidth
\def\findlength{\setlength\Linewidth\linewidth
	\addtolength\Linewidth{-4\fboxrule}
	\addtolength\Linewidth{-3\fboxsep}
}

\newenvironment{rqbox}{\par\begingroup
	\setlength{\fboxsep}{5pt}\findlength
	\setbox0=\vbox\bgroup\noindent
	\hsize=0.95\linewidth
	\begin{minipage}{0.95\linewidth}\normalsize}
	{\end{minipage}\egroup
	\textcolor{gray20}{\fboxsep1.5pt\fbox
		{\fboxsep5pt\colorbox{gray05}{\normalcolor\box0}}}
	\endgroup\par\noindent
	\normalcolor\ignorespacesafterend}

\AtBeginDocument{%
  \providecommand\BibTeX{{%
    \normalfont B\kern-0.5em{\scshape i\kern-0.25em b}\kern-0.8em\TeX}}}

\acmBooktitle{}

\begin{document}

\begin{titlepage}
  \input{preprint_frontpage.tex}
\end{titlepage}

\title{Detecting Security Fixes in Open-Source Repositories using Static Code Analyzers}

\author{Therese Fehrer}
\authornote{Therese Fehrer contributed to this research during her internship at SAP Security Research where she worked on her MSc thesis.}
\email{t.fehrer@uvt.nl}
\affiliation{
  \institution{JADS, Tilburg University}
  \city{'s-Hertogenbosch}
  \country{The Netherlands}
}

\author{Roc\'io Cabrera Lozoya}
\email{rocio.cabrera.lozoya@sap.com}
\affiliation{
  \institution{SAP Security Research}
  \country{France}
}

\author{Antonino Sabetta}
\email{antonino.sabetta@sap.com}
\orcid{0000-0003-3506-8374}
\affiliation{
  \institution{SAP Security Research}
  \country{France}
}

\author{Dario Di Nucci}
\email{d.dinucci@uvt.nl}
\orcid{10000-0002-3861-1902}
\affiliation{
  \institution{JADS, Tilburg University}
  \city{'s-Hertogenbosch}
  \country{The Netherlands}
}

\author{Damian A. Tamburri}
\email{d.a.tamburri@tue.nl}
\orcid{0000-0003-1230-8961}
\affiliation{
  \institution{JADS, Eindhoven University of Technology}
  \city{'s-Hertogenbosch}
  \country{The Netherlands}
}

\renewcommand{\shortauthors}{Fehrer, et al.}

\begin{abstract}
  \input{sections/abstract.tex}
\end{abstract}



\maketitle

\section{Introduction}
\input{sections/introduction.tex}

\section{Background}
\input{sections/background.tex}

\section{Empirical Study Goals and Data Preparation}
\input{sections/design.tex}

\section{Static Code Analyzers, Embeddings, and Security-relevant Commits}
\input{sections/rq1.tex}

\section{Levaraging on Static Code Analyzers Embeddings to Detect Security-relevant Commits}
\input{sections/rq2.tex}

\section{Threats to Validity}
\input{sections/threats.tex}

\section{Related Work}
\input{sections/related.tex}

\section{Conclusion and Future Work}
\input{sections/conclusion.tex}

\begin{acks}
  Antonino Sabetta is supported by the European Commission grant no. 952647 (AssureMOSS H2020).
  Dario Di Nucci and Damian A. Tamburri are supported by the European Commission grants no. 825040 (RADON H2020) and no. 825480 (SODALITE H2020).
\end{acks}


\bibliographystyle{ACM-Reference-Format}
\bibliography{main}

\end{document}

%% file: preprint_frontpage.tex


\enlargethispage{2\baselineskip}
\begin{center}
{\LARGE
{Detecting Security Fixes in Open-Source Repositories \\ using Static Code Analyzers}}\\[4mm]
{\large\bf [PRE-PRINT]} \\[6mm]
Therese Fehrer, Roc\'io Cabrera Lozoya, Antonino Sabetta, \\
Dario Di Nucci, Damian A. Tamburri
\end{center}

\thispagestyle{empty}
\vspace{8mm}
\small
\input{sections/abstract}
\vspace{5mm}
\normalsize
\hrule
\vspace{2mm}
\begin{center}
{\large Citing this paper}
\vspace{3mm}
\end{center}


\noindent
Please cite this work as:
\vspace{3mm}
{\tt
\begin{Verbatim}
@misc{fehrer2021detecting,
  title  = {Detecting Security Fixes in Open-Source Repositories
            using Static Code Analyzers},
  author = {Therese Fehrer and Roc\'io Cabrera Lozoya and Antonino Sabetta
            and Dario Di Nucci and Damian A. Tamburri },
  year   = {2021},
  month  = {March},
}
\end{Verbatim}
}

\vspace{3mm}
\hrule

\vspace*{\stretch{1}}
\includegraphics[width=4cm]{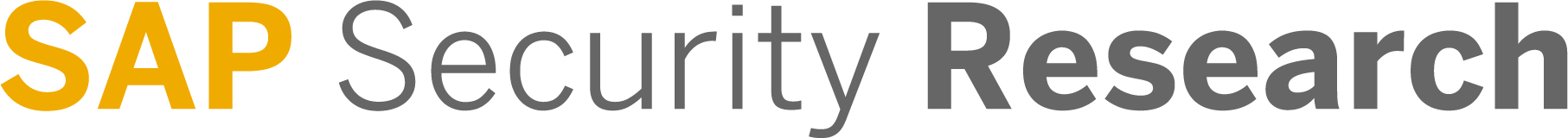}\hfill
\raisebox{-6mm}{\includegraphics[width=23mm]{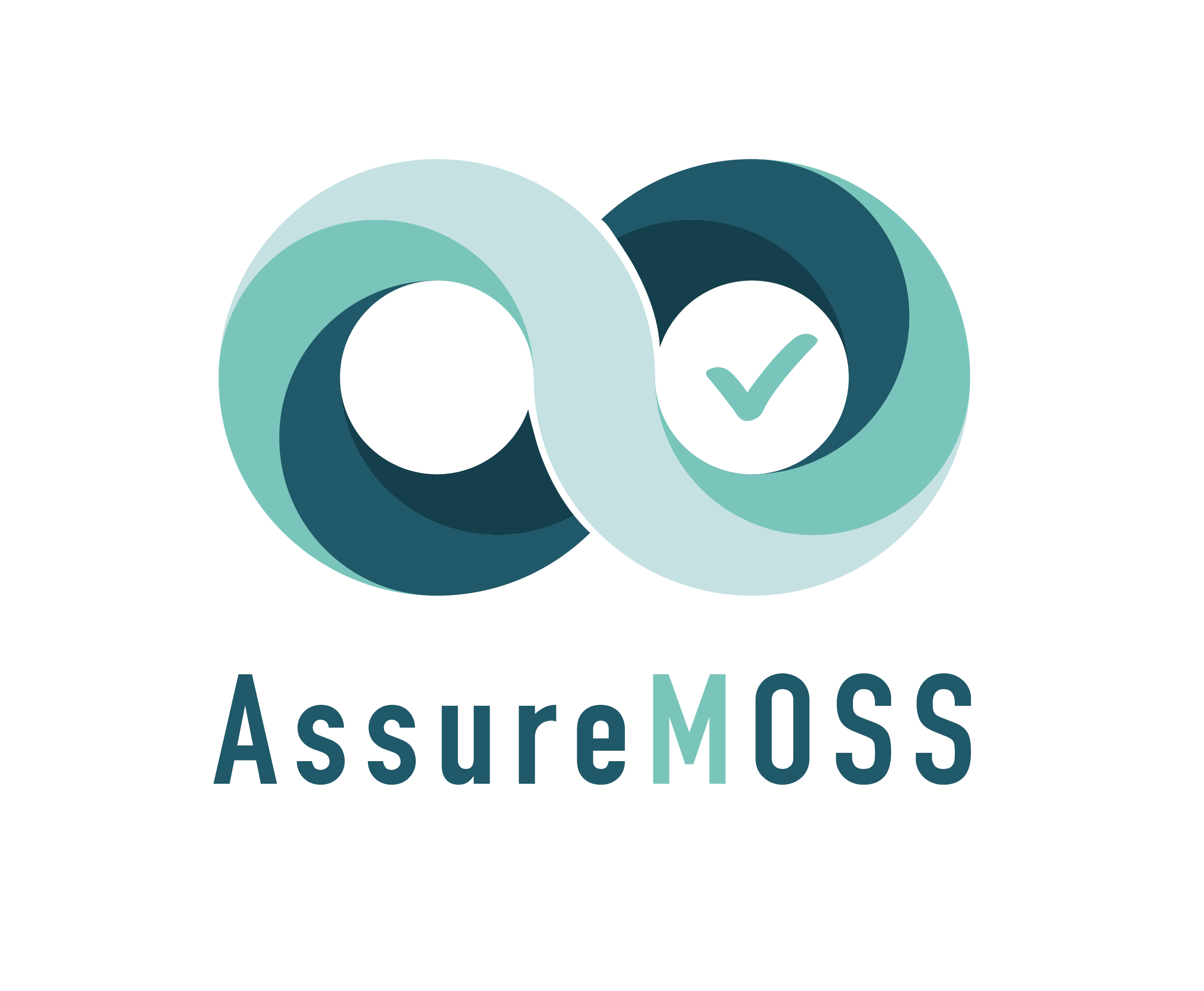}}\hfill
\raisebox{-2mm}{\includegraphics[width=18mm]{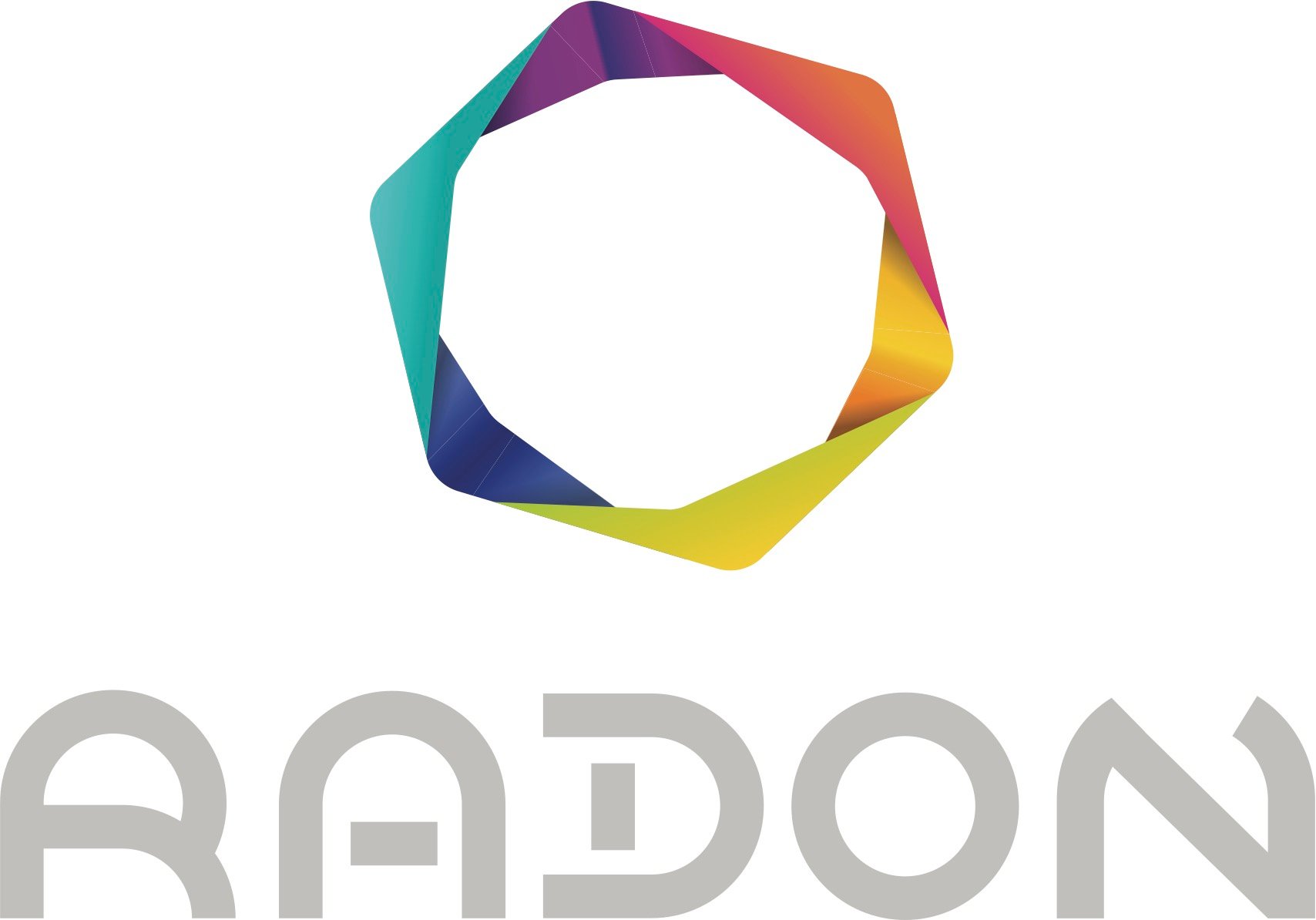}}\hfill
\raisebox{0mm}{\includegraphics[width=23mm]{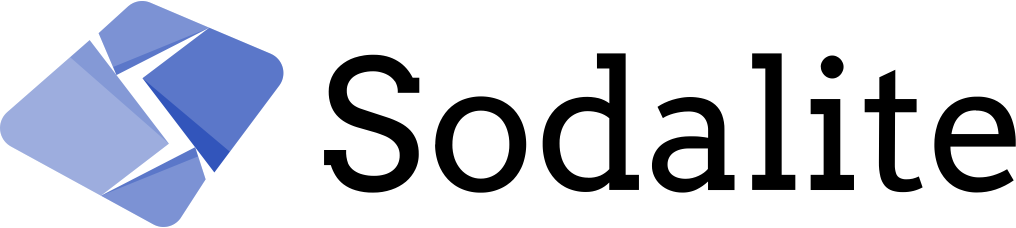}}


%% file: sections/abstract.tex

The sources of reliable, code-level information about vulnerabilities that
affect open-source software (OSS) are scarce, which hinders a broad adoption of
advanced tools that provide code-level detection and assessment of vulnerable
OSS dependencies.

In this paper, we study the extent to which the output of off-the-shelf static
code analyzers can be used as a source of features to represent commits in
Machine Learning (ML) applications. In particular, we investigate how such
features  can be used to construct embeddings and train ML models to
automatically identify source code commits that contain vulnerability fixes.

We analyze such embeddings for security-relevant and non-security-relevant
commits, and we show that, although \emph{in isolation} they are not different in a
statistically significant manner, it is possible to use them to construct a
ML pipeline that achieves results comparable with the state of the art.

We also found that the combination of our method with commit2vec represents a
tangible improvement over the state of the art in the automatic identification
of commits that fix vulnerabilities: the ML models we construct and commit2vec
are complementary, the former being more generally applicable, albeit not as
accurate.

%% file: sections/introduction.tex
The adoption of open-source software (OSS) components in commercial products has
dramatically increased over the past two decades. 
The large majority of commercial software products nowadays rely, to some
extent, on open-source software (OSS) components~\cite{ZaffarKZ11}. A study by
Snyk~\cite{tal_state_2019} reports that the large majority of the applications
they analyzed contained at least one open-source component; as much as 50\% to
the entire code-base of those applications is open-source.
Whereas traditionally software vendors used to have full control on the entire
development process of most of the components that made up their products,
nowadays a large part of the codebases of those products come from
community-developed free open-source projects, managed by independent parties~\cite{pashchenko2021assuremoss}.
By building upon these free, community-developed building blocks, vendors can focus their
efforts on differentiating features bring them to the market faster.
When doing so, however, they become responsible for assessing and mitigating the
impact that a vulnerability in those open-source components might have on their
products.

Advanced state-of-the-art tools such as \textsc{Eclipse
Steady}\footnote{\url{https://github.com/eclipse/steady}} use code-level
information about the vulnerable code fragment and its fix; this enables
efficient detection of vulnerable dependencies, as well as accurate impact
assessment through reachability analysis~\cite{ponta_emse_2020}. To do so,
Eclipse Steady relies on the availability of \emph{accurate data about the
code-level changes (commits) that fix each known vulnerability}.

Unfortunately, the data about security vulnerabilities is scattered across
heterogeneous sources, often not machine readable, and do not provide the
necessary level of detail, especially when it comes to code-level details.
The NVD would the natural candidate to provide such information, but as shown
in previous research~\cite{schryen_is_2011}
one cannot rely on NVD advisories to find fix commits.
Explicit mentions of CVE identifiers in commit messages are not frequent either
(in our dataset, only 12.18\% of the vulnerabilities have at least one fix commit
that can be found by searching for the CVE identifier in the commit messages).
Some attempts at curating vulnerability datasets do provide the necessary
level of details, but they suffer from scalability issues because they are
based on manual effort~\cite{ponta2019manually}.

This difficulty in obtaining accurate (code-level) vulnerability data hinders further development of
new tools that could push the state of the art in vulnerability detection and
mitigation. Also, it makes it harder for the research community to learn from
real-world vulnerability and their fixes.

In this perspective, the role of automated tools to find vulnerability fixes in code repositories,
possibly using machine learning, becomes increasingly
important~\cite{sabetta2018practical,lozoya2019commit2vec}. However, the research
that is necessary to come up with such tools requires, in turn, a large amount of training
data, which leads to a somewhat circular problem.



In this paper we present an approach to analyze repositories considering the
\emph{changes} introduced by individual commits. Instead of leveraging
specialized representations (such as syntax trees, call graphs, data flow
graphs, or their
combinations)~\cite{ghaffarian2017software,lozoya2019commit2vec}, we investigate
the use of off-the-shelf tools for code quality analysis to as sources of
features to represent code changes as captured in commits.  Based on
representations build from these features, we train models to predict whether a
commit is likely to fix a vulnerability or not. 
Static code analyzers have been combined with machine learning techniques for a
range of other software engineering-related prediction tasks. They have been
used in the context of software fault prediction~\cite{malhotra2015systematic,
li2018progress}, defect prediction~\cite{zimmermann2008predicting,
premraj2011network, tosun2009validation, bhattacharya2012graph}, code smell
prediction~\cite{lujan2020preliminary} and for the prediction and prioritization
of technical debt~\cite{codabux2016technical}.
We propose a model to represent code changes for whole repositories using static
code analyzers. Based on this representation, we then use machine learning
techniques to identify security-relevant commits and evaluate our model's
performance on an industrial dataset of vulnerabilities and their patches. The
analysis of the differences between security-relevant and non-security-relevant
commits reveals no statistically significant difference. Machine learning models
leveraging the results of static analysis tools achieve results comparable with
commit2vec. However, the latter is executable only on compilable code, being
based on Abstract Syntax Trees. Overall, a model combining both approaches
achieves better performance than those obtained by the approaches independently.

Our contributions are relevant from both an academic and an industrial
perspective. In particular, we provide:

\begin{itemize}
    \item new insights concerning the usage of \emph{off-the-shelf static code analyzer as sources of features}
     that can be used to predict security-relevant commits;
    \item an extensible pipeline on to pre-process and combine the outputs of
          various static code analyzers to train machine learning models on commits;
    \item a comprehensive empirical study that compares models trained on such features
          with alternative, state-of-the-art, methods that use information extracted
          from abstract syntax trees.
\end{itemize}

Our approach is relevant to industry as it is applicable to a much broader range of commits
than commit2vec~\cite{lozoya2019commit2vec}, the state-of-the-art tool
for this task, leveraging on the combination of different static code analyzers. Our method
can complement commit2vec yielding a tangible contribution to the goal of automating
the identification of \textit{security-relevant} commits (\ie commits that fix vulnerabilities).

\paragraph{Outline}
The remainder of this paper is structured as follows. \Cref{sec:background}
presents background information on software vulnerabilities and static code
analyzers. \Cref{sec:design} provides the goals of the empirical study and the
data preparation steps. \Cref{sec:rq1} describes how to extract embedding
resulting from the output of the static code analyzer. Then, it analyzes the
relationship between the embedding resulting from the output of the static code
analyzer and the security-relevant commit. \Cref{sec:rq2} introduces the
pipeline to train the prediction models and the method adopted to evaluate them.
It investigates and contrasts several machine-learning models and ensemble
techniques. Finally, it compares the obtained results with those achieved by
commit2vec. The threats to the validity of the empirical study are reported in
\Cref{sec:threats}, while \Cref{sec:related} discusses the related literature.
Finally, \Cref{sec:conclusion} concludes the paper and outlines the future work.

%% file: sections/background.tex
\label{sec:background}

The background of this research concerns software vulnerabilities and static analyzers, which we introduce in this section.

\subsection{Software Vulnerabilities}
Two of the most important institutions in the security ecosystem are the national vulnerability database (NVD) and Common Vulnerabilities and Exposures (CVE). The former is a U.S. government repository of vulnerability management data based on the CVE list and performs analysis on the vulnerabilities.\footnote{\url{https://nvd.nist.gov/general}} The latter is a list of known cybersecurity documented vulnerabilities, which include their identification numbers and descriptions.\footnote{\url{http://cve.mitre.org/about/}} It has become the standard for vulnerability and exposure identifiers. 

CVE defines a vulnerability as ``a flaw in software, firmware, hardware, or service component resulting from a weakness that can be exploited, causing a negative impact to the confidentiality, integrity, or availability of an impacted component or components''. In practice, different definitions of vulnerabilities are used, depending on the context and use case. This research originated from an industrial context through cooperation with SAP.  The definition of vulnerabilities is driven by practical considerations about the availability of relevant data sources. We follow the definition of vulnerability presented in commit2vec~\cite{lozoya2019commit2vec} since our research is based on the same industrial data set, where the vulnerabilities are implicitly defined through their associated fixes. 

Once a vulnerability is identified, a patch is created and made available to the public~\cite{frei2010modeling}. Software users need to install patches timely to minimize the time that they are under the risk of exploitation through the vulnerability. Commits that fix vulnerabilities are categorised as \textit{security-relevant commits}.

\subsection{Static Code Analyzers}
Static analyzers read a program and construct its abstract representation without executing it~\cite{louridas2006static} to analyze its quality. Static analyzers rely on \textit{pattern matching}, \textit{software metrics}, and \textit{program analysis}.
In the context of our research, we identified four static analyzers to generate embeddings for the commits in our dataset.

\paragraph{Static Code Analyzed based on Pattern Matching}
\textsc{PMD}\footnote{\url{https://pmd.github.io}} and
\textsc{Checkstyle}\footnote{\url{https://checkstyle.org}} are source code
analyzers designed to find common programming flaws and help to enforce coding
standards. Developers can use the standard rule sets provided or create their
own. In this way, the tools can be customized to fit different coding standards.
\textsc{PMD} currently supports eight different programming languages, including
Java and Apex. The standard ruleset for Java can detect eight bug categories.
\textsc{Checkstyle} is focused on Java code only and provides two configuration
files, which refer to the Sun Code conventions and the Google Java style, which
are available as predefined styleguides. Both \textsc{PMD} and
\textsc{Checkstyle} have been used previously used to detect code
smells~\cite{lujan2020preliminary}.
\citeauthor{lujan2020preliminary}~\cite{lujan2020preliminary} employs also
FindBugs\footnote{\url{http://findbugs.sourceforge.net}}. However, both FindBugs
and its successor SpotBugs\footnote{\url{https://spotbugs.github.io}} have the
limitation of working with bytecode rather than source code. Therefore, we did
not consider them for this research. There does, however, exist an extension for
SpotBugs, which is called Find Security
Bugs\footnote{\url{https://find-sec-bugs.github.io}} and is focused on
security-related warnings. Future research might want to evaluate if these
security-related bugs serve as a better feature set for training models to
identify security-relevant commits. 

\paragraph{Static Code Analyzed based on Software Metrics}
\textsc{CK} extracts class- and metric-level code metrics from Java source code.\footnote{\url{https://github.com/mauricioaniche/ck}} It calculates 30 different types of metrics, including five out of the six CK metrics~\cite{chidamber1994metrics}. The tool covers all CK metrics except the number of children (NOC), which was excluded due to memory issues. Please consider that the CK metrics have been used for various prediction tasks in software engineering, such as software fault prediction~\cite{malhotra2015systematic,di2017developer,li2018progress}.

\paragraph{Static Code Analyzed based on Program Analysis}
This category of static analyzers captures the inherent rich structure of source code.\textsc{Progex} ((Program Graph Extractor) extracts various graph representations like abstract syntax trees (ASTs), control flow graphs (CFGs), and program dependency graphs (PDGs) from source code.\footnote{\url{https://github.com/ghaffarian/progex}} It is based on the \textsc{ANTLR parser generator}\footnote{\url{https://antlr.org}} and currently supports only Java.

%% file: sections/design.tex
\label{sec:design}

The empirical study aims to understand to what extent machine learning
classifiers can leverage static code analyzers to automatically identify
security-relevant commits. These tools are widely used in the software
development process and create numerical representations of source code.
However, using such tools to create embeddings for commits has not been
sufficiently explored so far.

\begin{center}
        \begin{rqbox}
                \textbf{RQ1.} To what extent embedding resulting from static
                code analyzers can distinguish security-relevant commits?
        \end{rqbox}
\end{center}

One of the main challenges to apply machine techniques to detect
security-relevant commits is numerically representing them. Embedding commit
features means projecting the original feature space of the commits onto another
feature space, which should ideally capture the most relevant information of the
data points. Therefore, we aim at verifying to what extent static code analyzers
can be used to create embeddings for commits, which significantly distinguish
security-relevant commits.

\begin{center}
        \begin{rqbox}
                \textbf{RQ2.} To what extent machine learning models trained
                using embeddings resulted from static code analyzers identify
                security-relevant commits?
        \end{rqbox}
\end{center}

This research question analyzes the performance of the classification models
based on the previously defined embeddings.

To answer the research questions, we used a dataset of security-relevant
commits. We used a previous dataset~\cite{lozoya2019commit2vec} for several
reasons. First, the authors create embeddings for commits on a repository level
and use them to predict whether a commit is security-relevant or not. The
dataset allows us to benchmark our approach to commit2vec precisely. Second, the
dataset captures a representative sample of open-source projects of practical
industrial relevance. This aspect is essential as we want to evaluate the
usefulness of our approach in practice. Finally, part of the
dataset~\cite{ponta2019manually} is open-source, allowing future research to
replicate our study and assess the generality of our findings. 

\subsection{Data Collection}

We use the vulnerability knowledge base published by SAP in its \emph{project
``KB''} repository\footnote{\url{https://github.com/sap/project-kb}} and
described by Ponta et al.~\cite{ponta2019manually}. We extended this
dataset with additional records that were collected at SAP Security Research
after the publication of~\cite{ponta2019manually} and that are still
not published at the time of writing.

Based on the commits in the dataset thus obtained, we complemented it with
negative instances as explained
in~\cite{sabetta2018practical}: for each positive instance of a repository,
we extracted a random commit from the same repository, under the assumption that
security-relevant commits are relatively rare. Each commit obtained in this way was then
checked by looking for security-related keywords in the commit
message and manually inspected to exclude extremely large, empty, and
otherwise invalid commits. If these checks did not indicate that the commit is
invalid or security-relevant, it was added to the data set as a negative sample.
In other words, without explicit indications that a commit was
security-relevant, it was assumed that it was not security-relevant. 

The resulting dataset contains 1950 commits with an equal number of positive and
negative instances\footnote{While only part of this dataset is
public at this time, our study can be replicated using the content of the
dataset by Ponta et al~\cite{ponta2019manually}, which is public.}. 

\subsection{Data Pre-Processing}\label{data_preprocessing} We applied several
data pre-processing steps to the dataset~\cite{lozoya2019commit2vec}.
We ensured that only reachable commits were included in the data set and added
information concerning the vulnerability types; we removed
commits no longer reachable because of, for example, the deletion of the
respective branch. In particular, we excluded 133 commits (54 positive and 79
negative instances). Furthermore, repositories included in the data set only through
unreachable commits were dropped entirely. We aimed to keep the percentage of
positive and negative instances as close as possible to the original data set.
In the end, the final data set contained 1,821 commits, 129 less than the
original dataset.


%% file: sections/rq1.tex
\label{sec:rq1}

In this section, we answer the first research question which aims at
understanding to what extent the embeddings resulting from static code analyzers
can distinguish security-relevant commits.

\begin{figure*}[ht]
    \centering
    \includegraphics[width=0.70\textwidth]{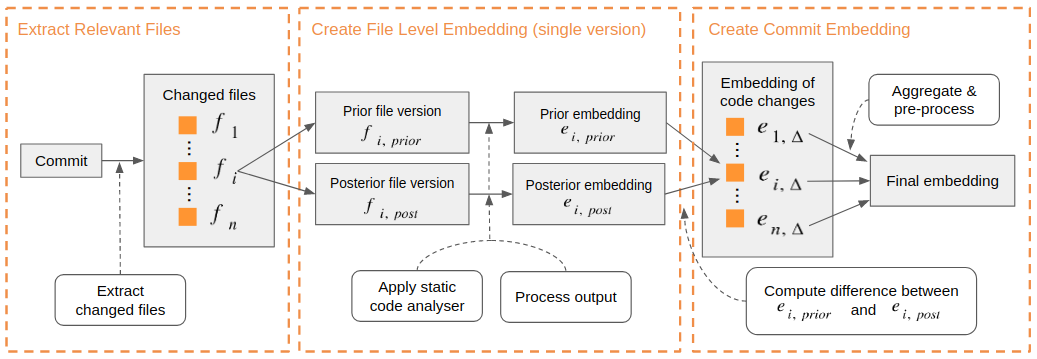}
    \Description{Embedding Pipeline}
    \caption{Embedding Pipeline}
    \label{fig:embedding_overview}
\end{figure*}

\subsection{RQ1 Study Design}

We extracted the embeddings for all commits in the dataset, as depicted in
\Cref{fig:embedding_overview}. We analyzed the files contained in the commits of
our dataset and their precessors using the static code analyzers introduced in
\Cref{sec:background}. The three main phases of the pipeline are (i) extracting
the relevant source code files, (ii) creating a numerical representation of
these files, and (iii) aggregating them into an embedding for the whole commit.
We computed one embedding per tool and applied several pre-processing steps to
obtain the final embeddings. Finally, we computed several statistical tests that
allowed us to answer RQ1.

\subsubsection{Extracting Relevant Source Code Files}
First, we identify which files of the repository were changed throughout the
commit. We collect the version before the commit (prior version) and the version
after the commit (posterior version) for each of these files. For performance
reasons, we only select files that are affected by a vulnerability in the
commit. Since we are interested in capturing code changes, files without any
change are not relevant to us. 

\subsubsection{Creating File Level Embeddings}
Second, we create embeddings for each of the collected files. We apply a static
code analyzer to each of the relevant files. Next, we create an embedding for
the files by transforming the tool's output into a numerical representation. The
specific transformation steps depend on the type of tool and will be explained
in the following. If a file is created or added, all measures for the
previous/current version are set to zero.

\paragraph{Static Code Analyzed based on Pattern Matching}
We use two different bug finders, \textsc{PMD} and \textsc{Checkstyle}, to
create embeddings that quantify bug changes between versions. Both tools return
the list of bugs and the lines where they occur. Based on this, we calculate how
often each bug occurs in the file. The result is a single version embedding with
one column per bug and one line per changed file and version. Each value
indicates how often a certain bug occurs in the file for a specific version.
Some bug descriptions in the output generated by \textsc{PMD} include
information like variable names. These bugs are renamed to more general names
without such case-specific information to compare them across different files
and repositories.  For PMD, we apply all rules from all categories available.
For \textsc{Checkstyle}, we used two example configuration files provided by the
tool.

\paragraph{Static Code Analyzed based on Software Metrics}
\textsc{CK} returns four CSV files: \texttt{class.csv}, \texttt{field.csv},
\texttt{method.csv}, and \texttt{variable.csv}. We only use the
\texttt{class.csv} file since it captures the software metrics at the file
level. Each line of this file represents one class in the repository and
contains the information concerning which file the class is located in, the
class type, different software metrics.

\paragraph{Static Code Analyzed based on Program Analysis}
To create this embedding, first, we extracted the graphs from the source code
files using Progex. Then, we measure their characteristics using several
metrics. We limit the analysis to ASTs and CFGs and analyze these graphs using
Networkx\footnote{\url{https://networkx.org/}}, a popular Python package for the
network analysis. Some measures can be computed if  only some condition is
fulfilled. In case these conditions are met, the score is set to zero for that
instance. For example, assortativity measures can be computed for graphs that
contain at least one edge since the similarity of connections can be computed if
such connections exist in the first place. Distance measures are meaningful only
for connected graphs such as ASTs. The number of components is informative only
for disconnected graphs such as CFGs because ASTs are always connected. 

\subsubsection{Computing Commit Embeddings}
We compute the file-level embedding of the commit. For each file, we compute the
difference between the measures in the previous and the current version. Next,
the file-level numerical representation is aggregated at the commit level. We
add two features indicating a positive or negative change for each feature at
the file-level embedding: the first to aggregate positive file-level changes and
the second to aggregate negative file-level changes. This step allows us to
avoid information loss: a simple sum over all the feature values could have
balanced positive and negative values. Finally, we drop all duplicate columns
and columns with constant values. 

\subsubsection{Evaluating Embedding based on Static Analyzer to distinguish Security-relevant Commits}
We use the Chi-square independence test to assess to what extent static code
analyzer can be employed to create embeddings to distinguish security relevant
commits (\ie RQ1). For each feature, the null hypotheses is that the feature is
independent (\ie does not impact) on the security relevance of a commit. We
choose Chi-square since it is a non-parametric test and, therefore, it is robust
to data distributions~\cite{mchugh2013chi}. Although most statistical tests
assume a random sample, this technique is commonly applied for convenience
samples. Chi-square is designed for two variables belonging to mutually
exclusive categories like in our case. Based on these variables, a contingency
table is created. The smallest number in any of the contingency table cells
should be no less than five~\cite{mchugh2013chi}. To comply with this
requirement, if more than 90\% of all values for a feature are zero, we convert
this feature into a binary variable. Otherwise, we create one class for all not
modified instances and bin the remaining values into $k$ different categories,
indicating the strength of the change in the variable. We select $k$ as the
highest number of possible equally-sized bins per feature to avoid splitting the
same values over multiple bins. In particular, we distributed the values of the
features resulting from the \textsc{PMD} and \textsc{Progex} over four bins; we
composed seven bins for each feature resulting from \textsc{CK} and five bins
for each feature resulting from \textsc{Checkstyle}. The remaining assumptions
are the following: (i) the data in the cells of the contingency table should be
counts, (ii) each subject (in our case: each commit) should only fall into one
cell of the table, and (iii) the classes should be independent. For contingency
tables with one degree of freedom, we used the recommended Yates' correction of
continuity~\cite{cochran1952chi2}. Since Chi-square is a test of significance
but not strength, we use Cramer's V when Chi-square rejects the null
hypothesis~\cite{mchugh2013chi}. Results below $0.1$ represent no association,
between $0.1$ and $0.3$ low association, between $0.3$ and $0.5$ moderate
association, and above $0.5$ high association. We implement Chi-square and
Cramer's tests in Python using the scipy library. We use Yates' correction and
interpret it with $\alpha = 0.05$.

\subsection{RQ1 Results}
We applied two statistical tests to evaluate whether there is a significant
difference between security-relevant and non-security-relevant commits for the
individual columns of the embeddings. First, we applied Chi-square; then, we
evaluated the strength of the association using Cramer's V. 

\paragraph{Chi-Square Test}

\begin{figure}[ht]
        \centering
        \includegraphics[width=\linewidth]{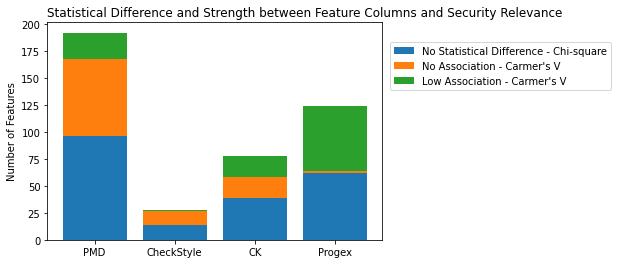}
        \Description{Features with significant code changes and their strength
        between security-relevant and non-security-relevant commits per
        embedding.}
        \caption{Features with significant code changes and their strength between security-relevant and non-security-relevant commits per embedding.}
        \label{fig:chi2_cramer}
\end{figure}

As depicted in \Cref{fig:chi2_cramer}, the embeddings based on \textsc{PMD} and
\textsc{Progex} have the highest number of features which show significant
differences between security-relevant and non-security-relevant commits with 96
and 62 features, respectively. If these features represent 84\% of features
resulting from \textsc{Progex}, they represent only 21\% of features for
\textsc{PMD}. 45\% of all features in \textsc{CK} show significant differences
between classes, while only 11\% of features (i.e., 14 out of 122) resulting
from Checkstyle significantly capture differences.

\paragraph{Cramer's V} 
For the features that showed a significant association with the target variable,
we computed Cramer's V to assess the strength of this association.
\Cref{fig:chi2_cramer} shows the results for this test. All features of the four
embeddings have a low association or no association at all. Almost all features
of the \textsc{Progex} embedding, which showed significant differences between
classes, had a low association between the target variable and the respective
column. For Checkstyle, on the other hand, all columns except one showed no
association in terms of Cramer's V. 

\begin{center}
\begin{rqbox}
    \textbf{Summary of RQ1.} Taken alone, most of the features resulting from
    the static analysis tools are not significantly different for
    security-relevant and non-security-relevant commits. If differences are
    significant, the association strength is low. 
\end{rqbox}
\end{center}

%% file: sections/rq2.tex
\label{sec:rq2}

In this section, we answer the second research question which aims at
understanding to what extent the embeddings resulting from static code analyzers
can be used to train machine learning models to classify security-relevant
commits.

\begin{figure*}[ht]
    \centering
    \includegraphics[width=0.70\textwidth]{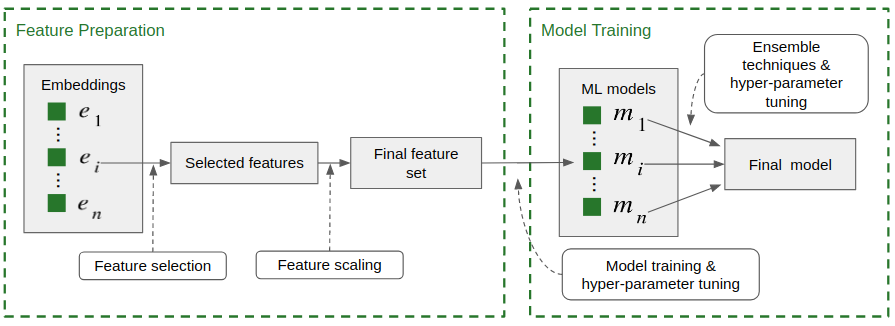}
    \Description{Machine Learning Pipeline}
    \caption{Machine Learning Pipeline}
    \label{fig:ml_overview}
\end{figure*}

\subsection{RQ2 Study Design}

We trained multiple prediction models to identify security-relevant commits
using the embeddings created in the previous section. \Cref{fig:ml_overview}
depicts an overview of the adopted machine learning pipeline. First, we train
models separately for each of the four embeddings. The embeddings are
pre-processed using feature selection and feature scaling. We train a model per
embedding using seven different machine learning algorithms and tune the
hyperparameter values of the feature pre-processing and model training steps
through a randomized search.

\paragraph{Dependent Variable}
The variable to be predicted is the security relevance of a commit, where one
indicates that a commit fixes a vulnerability and zero that it does not. The
prediction problem is a binary classification problem.

\paragraph{Independent Variables}
The independent features are the previously described embeddings, computed using
the four static analysis tools. In the first step, the embeddings are used
separately (\ie one model per embedding). Then, these models are combined using
voting and stacking techniques.

\paragraph{Evaluation Metrics}
Performance metrics of binary classification problems are inherently tied to the
number of correct and incorrect predictions for both classes, which are often
visualized in a confusion matrix. The focus of this research is precision, which
is interpreted along with recall using precision-recall curves. PR curves are a
useful alternative to ROC curves to highlight the performance differences that
are lost in ROC curves~\cite{boyd2013area}. Furthermore, we also calculate
accuracy and F1 scores~\cite{baeza1999modern} to capture the general performance
of our models.

\paragraph{Model Validation}
Each model is trained and tuned using stratified k-fold cross-validation with
$k=5$. $k$-fold cross-validation randomly splits the dataset into $k$ mutually
exclusive subsets of equal size. The predictor is then trained $k$ times, each
time using one fold for testing and the remaining $k-1$ folds for training. In
stratified cross-validation, the folds contain approximately the same proportion
of labels as the original dataset~\cite{kohavi1995study}. We use the same folds
used in commit2vec~\cite{lozoya2019commit2vec} leveraging on the "test\_fold"
variable in the dataset that indicates the fold to which each commit belongs.

\subsubsection{Feature Selection}
Feature selection helps to reduce computation time, improve prediction
performance, and give a better understanding of the data by removing correlated
variables, which do not provide extra information about the classes and thus
serve as noise for the predictor. We apply three different types of feature
selection techniques. First, we remove features whose variance falls below the
chosen threshold~\cite{li2017feature}. This step is especially relevant for the
static analyzers based on pattern matching, which are very sparse. Features with
low variance contain little information and can, therefore, not discriminate
between classes. Since a feature's variance depends on the scale, we applied a
min-max
scaler\footnote{\url{https://scikit-learn.org/stable/modules/generated/sklearn.preprocessing.MinMaxScaler.html}}
before calculating the features' variances. We only use scaled features only for
the selection: the final feature set consists of the unscaled version of these
features. Second, we apply the Pearson correlation
coefficient~\cite{pearson1895vii}. Third, we perform recursive feature
elimination~\citeauthor{guyon2002gene} \cite{guyon2002gene}. This technique is
an example of backward feature elimination, starting with the full set of
features and removing features iteratively. Each iteration consists of three
steps: (i) train the classifier, (ii) compute the ranking criterion for all
(remaining) features, (iii) remove the feature (or the set of features) with the
smallest ranking criterion. We implement RFE using scikit-learn
\footnote{\url{https://scikit-learn.org/stable/modules/generated/sklearn.feature_selection.RFE.html}}.
We use these techniques independently and together, applying random search to
find the best combination of techniques and the best hyperparameters.
The parameter grid for feature selection can be found in the online appendix\footnote{\url{https://github.com/dardin88/fse_2021/blob/main/feature_selection_tuning.md}}.

\subsubsection{Feature Scaling}
To prepare the predictive variables for model training,  we apply scikit-learn's
StandardScaler\footnote{\url{https://scikit-learn.org/stable/modules/generated/sklearn.preprocessing.StandardScaler.html}},
which scales and centers features by subtracting the mean and dividing by the
standard deviation. We scale the feature set since some machine learning
techniques can be negatively impacted by data distribution and scale differences
between features.

\subsubsection{Base Machine Learning Classifiers}
First, we train multiple different models for each of the four embeddings.
Specifically, we use seven different algorithms for model training:
\textsc{Decision Trees}, \textsc{Random Forests}, \textsc{AdaBoost},
\textsc{Gradient Boosting}, \textsc{Support Vector Machines}, \textsc{Logistic
    Regression}, and \textsc{Gaussian Na\"ive Bayes}. These classifiers belong to
different families and complement each other in terms of pros and cons, mixing
some of the simplest techniques (\ie \textsc{Gaussian Na\"ive Bayes} and
\textsc{Logistic Regression}), with more complex ones (\ie \textsc{Random
    Forests}, \textsc{AdaBoost}, and \textsc{Gradient Boosting}). All of the above
techniques were implemented using
scikit-learn\footnote{\url{https://scikit-learn.org/stable/supervised_learning.html}},
a well-known Python framework.

\paragraph{Hyperparameter Tuning}
For each embedding, we train the models using the machine learning techniques
described before. We evaluate the effect of feature selection by training each
model once with and without feature selection. Please consider that we could not
apply Recursive Feature Elimination to  Gaussian Na\"ive Bayes and Support Vector
Machines since they do not provide feature importances. A random search over the
hyperparameter space is applied using scikit-learn's
RandomizedSearchCV\footnote{\url{https://scikit-learn.org/stable/modules/generated/sklearn.model_selection.RandomizedSearchCV.html}}.
This algorithm randomly samples the hyperparameter from the distributions
specified as part of the online appendix\footnote{\url{https://github.com/dardin88/fse_2021/blob/main/base_classifiers_tuning.md}}, those not specified
are kept to their default values provided by scikit-learn. We run the randomized
search for 200 iterations per machine learning technique, embedding, and
combination of feature pre-processings steps. We selected this method instead of
a more refined (\ie GridSearch) because the base models are optimized on many
hyper-parameters. Some are tuned along with numerical scales leading to the
high-dimensional optimization problem.

\subsubsection{Ensemble Techniques}
In a second step, the models trained on the separate embeddings are combined
through ensemble techniques, which combine multiple different classifiers to
make a final prediction. We select the best performing model for each embedding
regarding feature selection, machine learning algorithm, and hyper-parameter
values. We use Voting and Stacking to combine these base models.
Voting~\cite{kittler1998combining} combines several estimators either by taking
a majority vote or by averaging the predicted probabilities of the base
estimators to make a final prediction. We use the VotingClassifier provided by
scikit-learn\footnote{\url{https://scikit-learn.org/stable/modules/ensemble.html\#voting-classifier}}.
In particular, we use the soft vote option, which refers to the second case of
averaging the predicted probabilities. Stacking~\cite{wolpert1992stacked} stacks
the predictions of the individual estimators together and uses them as input to
a final estimator to compute the prediction. The final estimator is trained
through cross-validation. We use the StackingClassifier provided by
scikit-learn\footnote{\url{https://scikit-learn.org/stable/modules/ensemble.html\#stacked-generalization}}.

\paragraph{Hyperparameter Tuning}
The hyper-parameter settings of the final classifiers are optimized for
precision using
GridSearchCV\footnote{\url{https://scikit-learn.org/stable/modules/generated/sklearn.model_selection.GridSearchCV.html}}
on the same five-folds as in the base models' case. We adopted this technique
because of the smaller number of hyperparameters. Therefore, we want to test all
possible combinations in the hyperparameter space. The parameter grid for each
ensemble technique can be found in the online appendix\footnote{\url{https://github.com/dardin88/fse_2021/blob/main/ensemble_techniques_tuning.md}}. The
hyperparameters, which are not featured in the tables, were left at their
default values. We set up the weights of the base models for Voting to either
fully include or exclude each base model in the prediction. This step allows
examining which models complement each other and, following, which combination
of models results in the best predictions. When using Stacking, we employed all
previously described base models as final estimators.

We analyze the performance of the machine learning models, which were trained on
the embeddings. We discuss the quality of the models trained first with base
classifiers and then applying stacking and voting. We compare the results for
these two different ensemble techniques and discuss whether the ensemble models
reach higher predictive performance than the base models. Finally, we compare
our best model and commit2vec in terms of predictive performance and training
time and discuss the extent to which the two approaches are complementary.

\subsection{RQ2 Results}

We provide the results for the base machine learning classifiers, the ensemble
techniques, and the comparison with commit2vec.

\begin{figure}[ht]
    \centering
    \includegraphics[width=\linewidth]{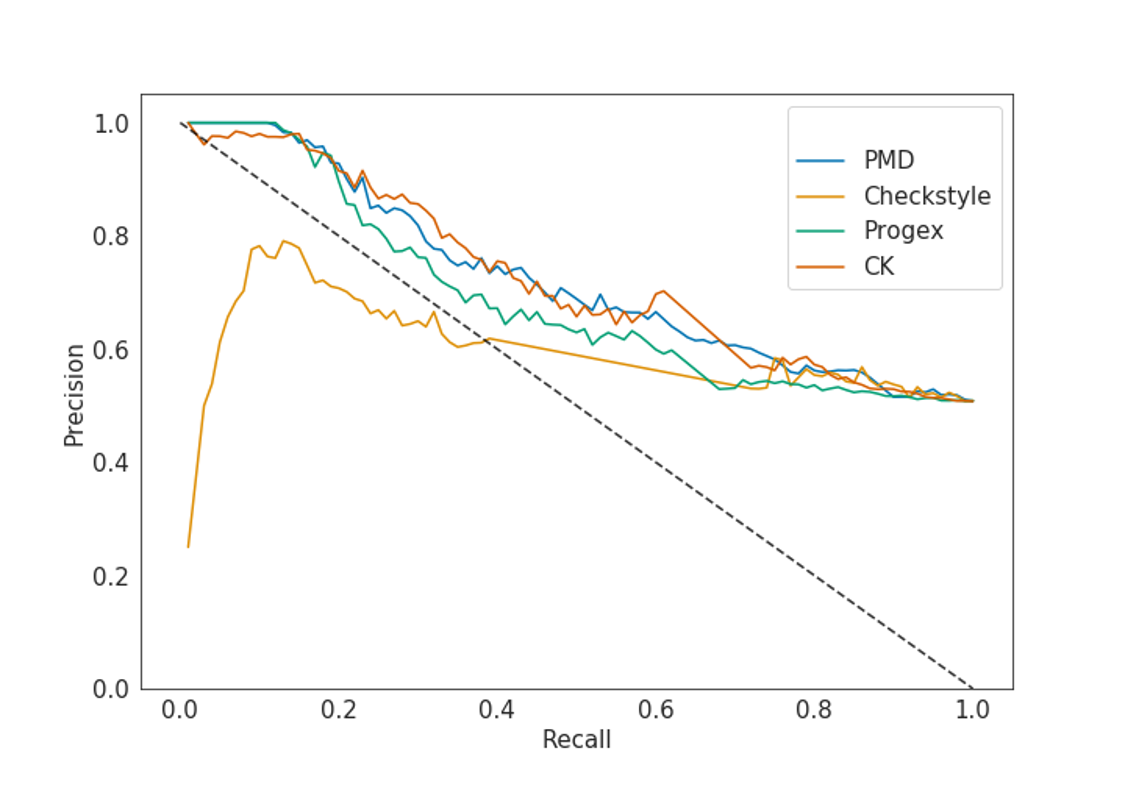}
    \Description{Precision-recall curves of the best model per embedding. The
        curves have been averaged over the five folds.}
    \caption{Precision-recall curves of the best model per embedding. The curves have been averaged over the five folds.}
    \label{fig:precision_recall_single}
\end{figure}

\input{tables/single_models.tex}

\subsubsection{Base Machine Learning Classifiers}
\Cref{fig:precision_recall_single} presents the precision-recall curves for the
best performing model of each embedding. As can be seen, the models based on
\textsc{PMD} and \text{CK} perform the best. The model trained on the
\textsc{Progex} embedding reached a good but lower performance. Although the
\textsc{CK} model is close to the other models' performance for thresholds with
high recall and low precision, it does not perform well when the target is high
precision.

\Cref{tab:single_models} show the performance of the best models for each
embedding in terms of precision, recall, accuracy, and F1 score when optimizing
the prediction threshold for precision. For the specific thresholds chosen, the
\textsc{CK} model performs the best regarding all four evaluation metrics.
Similarly, the \textsc{Checkstyle} model reaches the worst performance for all
four metrics. The \textsc{PMD} model performs better than the \textsc{Progex}
model regarding precision and accuracy but performs worse in terms of recall and
F1 score.

Analyzing the best models in details, we can note that the models based on
\textsc{PMD} and \textsc{CK} adopt \textsc{Random Forest}, while those
leveraging the features extracted from \textsc{CheckStyle} and \textsc{Progex}
use \textsc{Gradient Boosting}. For the sake of space, we report the
configuration of these models in our online appendix\footnote{\url{https://github.com/dardin88/fse_2021/blob/main/best_models.md}}.

\begin{figure}[ht]
    \centering
    \includegraphics[width=\linewidth]{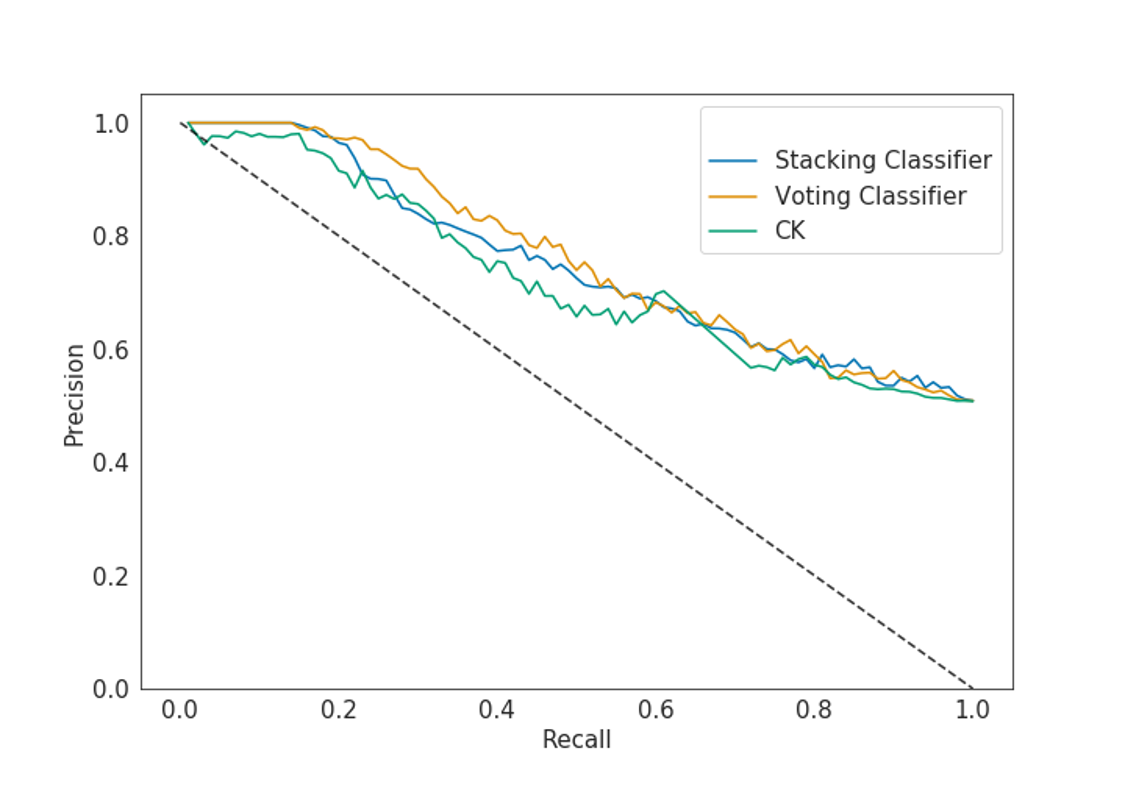}
    \Description{Precision-recall curves of the best model per ensemble. The
        curves have been averaged over the five folds.}
    \caption{Precision-recall curves of the best model per ensemble. The curves have been averaged over the five folds.}
    \label{fig:precision_recall_ensemble}
\end{figure}

\input{tables/ensemble_models}

\subsubsection{Ensemble Techniques}
We combined the four models from the previous section using the two different
ensemble techniques: stacking and voting. \Cref{fig:precision_recall_ensemble}
shows the precision-recall curve for both the final models. It can be seen that
for thresholds with high recall and lower precision, the two models perform very
similarly. For our aim of maximizing precision, however, the voting classifier
performs better. We then optimized the prediction thresholds of both models in
terms of precision. \Cref{tab:ensemble_models} show the values of the evaluation
metrics for the chosen thresholds. The voting classifier performs better than
the stacking classifier in terms of all four metrics and achieves an average
precision of 77.5\% and an average recall of 48.64\%.

We compared the precision-recall curves of the voting classifier to those of the
models based on the individual embeddings. Combining the models leads to
significantly better results, indicating that the models are complementary to
each other.

\subsubsection{Comparison with commit2vec}

\begin{figure}[!ht]
    \centering
    \includegraphics[width=\linewidth]{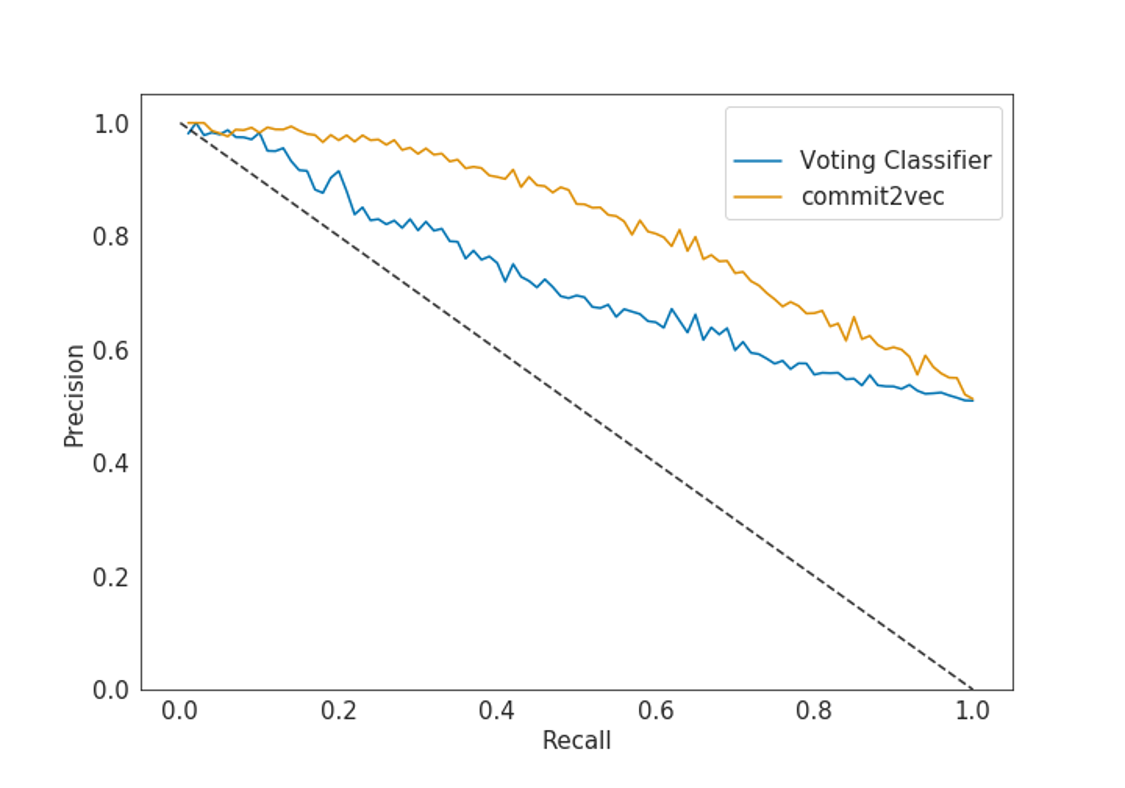}
    \Description{Precision-recall curves of commit2vec and voting classifier on
        the commits analyzable by commit2vec. The curves have been averaged over
        the five folds.} \caption{Precision-recall curves of commit2vec and
        voting classifier on the commits analyzable by commit2vec (\ie $\approx
        20\%$ of commits in our dataset). The curves have been averaged over the
        five folds.}
    \label{fig:precision_recall_commit2vec_vs_ensemble}
\end{figure}

Applying Commit2vec imposes several technical constraints, mostly stemming from
implementation limitations, some of which inherited from code2vec (on which
commit2vec is based). In particular, code2vec represents \emph{methods}, and
based on this represention, Commit2vec represents  commits in terms of
\emph{method changes}. As a consequence, that commits that do not change methods
cannot be processed. Furthermore, when analyzing method changes, commit2vec
works under the condition that the name of methods is not altered by a commit,
which of course is not always the case\footnote{For the same reason, methods
that are added from scratch or removed by a commit, cannot be represented by
commit2vec.}.

Therefore, commit2vec could process only around 20\% of the commits in
our dataset. However, as shown in
\Cref{fig:precision_recall_commit2vec_vs_ensemble}, commit2vec performs better
than our models on the commits that it could process. For prediction thresholds
resulting in high precision and lower recall, the performance of the two models
is very similar. However, for thresholds with lower precision, commit2vec
performs better.

\begin{figure}[!ht]
    \centering
    \includegraphics[width=\linewidth]{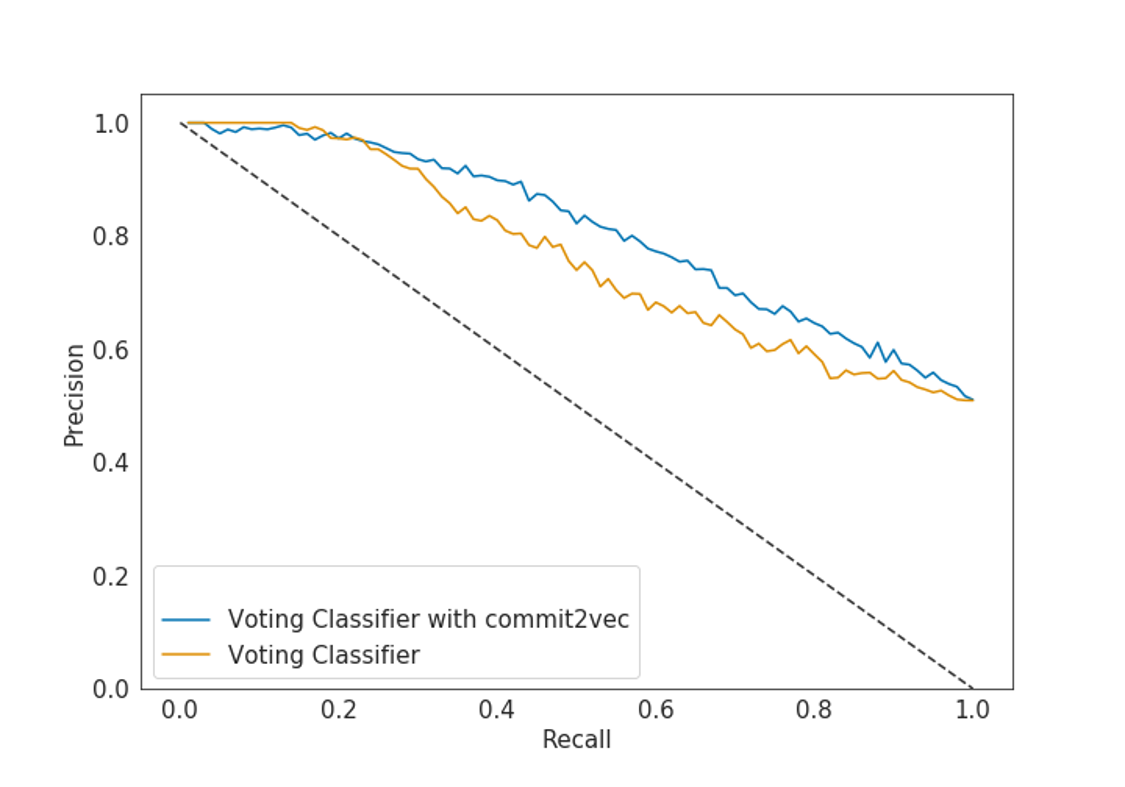}
    \Description{Precision-recall curves of the voting classifier and the
        combination of the voting classifier and commit2vec. The curves have
        been averaged over the five folds.}
    \caption{Precision-recall curves of the voting classifier and the combination of the voting classifier and commit2vec. The curves have been averaged over the five folds.}
    \label{fig:precision_recall_commit2vec_and_ensemble}
\end{figure}

In practice, our method and commit2vec turn out to be complementary: one could
use commit2vec when possible, and fall-back to our models in the other cases.
\Cref{fig:precision_recall_commit2vec_and_ensemble} visualizes the
precision-recall curves of the voting classifier and the model obtained in the
previously described way. Combining the two models, we can achieve much better
performance than using any of them independently.

An in-depth comparison of the training times for commit2vec and our approach is
out of the scope of this research. However, please note that our approach is
likely to be much faster. Complex deep learning approaches like commit2vec are
typically computationally very expensive. Code2vec, on which commit2vec is
based, takes 1.5 days on a single Tesla K80 GPU to thoroughly train a
model~\cite{alon2019code2vec}. In comparison, our approach takes less than half
a day on an ordinary laptop, giving a strong indication that our approach is
much less computationally expensive than commit2vec.

\begin{center}
    \begin{rqbox}
        \textbf{Summary of RQ2.}
        Machine learning models constructed from features obtained from static
        analysis tools yield a slighly lower prediction performance than
        commit2vec. However, they can work with commits that commit2vec is not
        capable of representing. Ultimately, the two approaches complement each
        other and are best used in combination.
    \end{rqbox}
\end{center}

%% file: tables/single_models.tex
\begin{center}
\begin{table}[ht]
\caption{The evaluation metrics of the best performing model per embedding. The table displays the mean and standard deviations of the metrics for each embedding.}
\label{tab:single_models}

\resizebox{0.48\textwidth}{!}{
\begin{tabular}{l|c|c|c|c}
\hline 
\textbf{Embedding} & \textbf{Precision} & \textbf{Recall} & \textbf{F1-Score} & \textbf{Accuracy}\\ \hline 
CK & 75.62 $\pm$ 5.60 &40.07 $\pm$ 1.28 &52.34 $\pm$ 2.11 & 63.05 $\pm$ 2.50 \\
Checkstyle & 64.89 $\pm$ 3.69 &31.25 $\pm$ 3.36 &42.05 $\pm$ 2.78 &56.56 $\pm$ 1.44 \\
PMD & 75.44 $\pm$ 5.37 & 35.50 $\pm$ 4.01 &48.05 $\pm$ 3.30 &61.35 $\pm$ 1.58 \\
Progex & 69.27 $\pm$ 4.01 &38.25 $\pm$ 4.14 &49.23 $\pm$ 4.22 &60.18 $\pm$ 3.15 \\
\hline 
\end{tabular}}

\end{table}
\end{center}

%% file: tables/ensemble_models.tex
\begin{center}
\begin{table}[ht]
\caption{The evaluation metrics of the best performing model per ensemble technique. The table displays the mean and standard deviations of the metrics for each ensemble technique.}
\label{tab:ensemble_models}

\resizebox{0.48\textwidth}{!}{
\begin{tabular}{l|c|c|c|c}
\hline 
\textbf{Embedding} & \textbf{Precision} & \textbf{Recall} & \textbf{F1-Score} & \textbf{Accuracy}\\ \hline 
Stacking & 74.75 $\pm$ 4.87 &45.89 $\pm$ 5.02 &56.56 $\pm$ 2.94 & 64.52 $\pm$ 1.60 \\
Voting & 77.51 $\pm$ 5.28 &48.64 $\pm$ 1.71 &59.68 $\pm$ 1.55 &66.73 $\pm$ 2.13 \\
\hline 
\end{tabular}}

\end{table}
\end{center}

%% file: sections/threats.tex
\label{sec:threats}
This section describes the threats to validity of our study. 

\paragraph{Construct Validity}
This research is inherently based on the definition of software vulnerabilities
and commits that are likely to fix such vulnerabilities. The classification of
commits in our dataset into security-relevant and non-security-relevant commits
is not based on a formal definition. The dataset was created based on practical
considerations, and vulnerabilities are only implicitly defined through their
fix-commits. Therefore, our data could not fully represent vulnerabilities and
their fix-commits. 

\paragraph{Internal Validity}
The results of our study are tied to the specific static code analyzers and
pre-processing steps of their outputs, which we used to create the embeddings.
Other static code analyzers or other pre-processing approaches may significantly
change the results. We mitigated this risk through researching, which static
code analyzers and pre-processing steps have been applied in related software
engineering prediction tasks and following a similar research design. 

\paragraph{External Validity}
The dataset we used contains repositories, which are practical relevant to SAP,
the company which provided the dataset. Within these repositories, there might
be a bias towards easily reachable commits. Since part of the dataset was
manually curated, certain types of commits might have been harder to reach and
not represented appropriately in the dataset. Furthermore, our data distribution
might not reflect the actual distribution of vulnerability types. Due to the
reasons above, the generalizability of our findings to other industrial or
academic settings might be limited. 

\paragraph{Conclusion Validity}
The data, which we used for the statistical test to address RQ2 does not fully
comply with the assumptions of the chi-square test. First, the data was not
obtained through random sampling. Second, not all cells in the contingency table
initially had values of five or higher, which might affect the results of the
test. We mitigated this risk by binning the initial data into several categories
to ensure that all cells in the contingency table have values of at least five.
The metrics used to evaluate our defect prediction approach (\ie precision,
recall, F-Measure, and accuracy) are widely used in the evaluation of the
performances of defect prediction
techniques~\cite{boyd2013area,baeza1999modern}. We used the precision-recall
curves, an alternative and more conservative visualization than ROC curves, to
evaluate the models' overall performance.

%% file: sections/related.tex
\label{sec:related}

This section provides a comprehensive overview of the previous work related to
vulnerability prediction and methods levering on features resulting from static
analysis tools.

\subsection{Vulnerability Prediction}
Several surveys exist on the mitigation of software vulnerabilities.
\citeauthor{shahriar2012mitigating}~\cite{shahriar2012mitigating} present and
categorize work on the topic, published between 1994 and 2010. They found that
most approaches fall into the following three categories: dynamic analysis,
static analysis, and hybrid analysis. A later survey by
\citeauthor{ghaffarian2017software}~\cite{ghaffarian2017software} covers the
work between 2000 and 2016 and focuses on the application of machine learning
and data mining approaches to analyze software vulnerabilities. Common
approaches include vulnerability prediction models based on software metrics,
anomaly detection approaches, and vulnerable code pattern recognition.  
The performance of vulnerability analysis models is typically analyzed in terms
of false positives and false negatives. Both reviews report that no solution
that is satisfactory on both dimensions has been found so far. In practice, the
main challenge remains the false positive rate, as it requires large amounts of
manual labor to identify vulnerable code instances among an extensive set of
positive predictions. Existing research deviates from this work in multiple
ways; the most crucial difference is the unit of analysis. In this work, we are
not interested in identifying if a software component is vulnerable but instead
if a commit is likely to fix a vulnerability. Furthermore, the focus of most
existing approaches is a method or file-level analysis. We instead use whole
repositories as our unit for analysis.

The closest work to ours is commit2vec~\cite{lozoya2019commit2vec}. Commit2vec
represents whole commits on a repository level to identify security-relevant
commits. The model is based on code2vec~\cite{alon2019code2vec}, which creates
continuous embeddings for code snippets relying on abstract syntax trees and
neural networks. Because of the former, Commit2vec can detect vulnerability only
on compilable classes. Our approach does not exhibit this limitation, as we are
using a broader set of features to build our models, which is not restricted to
features extracted from the AST. Next to program analysis-related features, we
also create features based on pattern matching and software metrics.

\citeauthor{nguyen2010predicting}~\cite{nguyen2010predicting} compute code
metrics based on dependency graphs, obtained by static code analyzers, to
predict vulnerable software components. This study computed a set of metrics on
the dependency graph of software systems and used them as input for machine
learning models. The empirical results were promising and suggested the
effectiveness of using dependency graphs for detecting software vulnerabilities.
Our work focus on a different unit of analysis, focusing on code changes instead
of single versions. Furthermore, our work examines the usefulness of a broader
range of static code analyzers for vulnerability prediction. 

\subsection{On the Use of Static Code Analysers in Software Engineering Prediction Tasks}
Static code analyzers have been used to create different feature sets for
prediction tasks in software engineering: software metrics, graph-based metrics,
and patter-matching-based metrics. 

\subsubsection{Pattern-Matching Metrics}

\citeauthor{lujan2020preliminary}~\cite{lujan2020preliminary} recently used the
output of bug finders to predict code smells, namely poor implementation choices
applied during software evolution that can affect source code
maintainability~\cite{fowler2018refactoring}. As more attention has been
dedicated to machine learning solutions for identifying code
smells~\cite{di2018detecting}, the authors examine the role of static analysis
warnings as features of machine learning models for detecting code smell types.
Specifically, \textsc{Checkstyle}, \textsc{FindBugs}, and \textsc{PMD} are used
to detect three types of code smells. Based on the output of such tools, a
feature set was constructed by providing the number of violations exhibited by
the software components for each warning. The evaluation results show that the
warning types that contribute the most to the performance of the models depend
on the considered code smell. Furthermore, the performance of the models built
using the warnings of static analysis tools can drastically improve the
capabilities of alternative code smell prediction models.

\subsubsection{Software Metrics}
Software metrics have also been used as a feature set for a variety of other
prediction tasks. Most notably, they have been used in software fault and
technical debt prediction. 

Software fault prediction aims to detect faulty constructs such as modules or
classes in the early phases of the software development life
cycle~\cite{malhotra2015systematic,li2018progress}.
\citeauthor{malhotra2015systematic}~\cite{malhotra2015systematic} reports that
most studies use software metrics as independent variables for the prediction
task. The mainly usedsoftware metrics  are procedural, object-oriented, or
hybrid metrics.  CK metrics, defined by by Chidamber and
Kemerer~\cite{chidamber1994metrics}, are the most commonly used software
metrics, out of which multiple metrics were found helpful for software fault
prediction. The survey results indicate that machine learning techniques perform
better than traditional statistical models for software fault prediction.
\citeauthor{li2018progress}~\cite{li2018progress} divide software metrics into
code and process metrics. Code metrics represent the source code complexity,
while process metrics represent the development process complexity. They also
found that the CK metrics are among the most representative metrics for
object-oriented programs. 

Technical debt refers to design decisions that allow software development teams
to achieve short-term benefits such as timely code releases. Technical debt
needs to be managed carefully to avoid severe consequences in the future. The
technical debt items are usually documented, including the required effort to
resolve the item. Then, the items are prioritized and addressed accordingly.
\citeauthor{codabux2016technical}~\cite{codabux2016technical} propose a
framework to prioritize technical debt items using data mining and machine
learning models. The authors extract class-level software metrics for defect-
and change-prone classes. Then, a prediction model is built based on these
metrics to determine the technical debt proneness of each class. The predictions
are finally used to categorize the items according to their debt proneness
probability. 

\subsubsection{Graph-Based Metrics}
Several papers use dependency graphs to identify program units that are more
likely to face defects.
\citeauthor{zimmermann2008predicting}~\cite{zimmermann2008predicting} propose
using network analysis on the dependency graphs to train regression models to
predict defects. The results were deemed promising as the graph-based models
performed significantly better than the code metrics-based models. Two
replication studies were conducted to test the generality of the findings with
contrasting results~\cite{premraj2011network,tosun2009validation}.
\citeauthor{bhattacharya2012graph}~\cite{bhattacharya2012graph} considered a
broader scope by using network metrics for bug severity estimation and
refactoring efforts prioritization, next to the prediction of defects. Their
approach is based on graphs at different levels of granularity. The evaluation
of this approach, conducted on multiple programs, indicates that graph metrics
can predict bug severity, maintenance effort, and defect-prone releases.

%% file: sections/conclusion.tex
\label{sec:conclusion}

In recent years, the need for automated identification of security-relevant
commits emerged to ensure software security. We address this need by proposing a
framework to use static code analyzers combined with machine learning techniques
to detect security-relevant commits.  We created four different embeddings based on
off-the-shelf code analyzers (\textsc{PMD}, \textsc{Checkstyle}, \textsc{CK}, and \textsc{Progex}).
We statistically evaluated whether the features of the embeddings differ
significantly for security-relevant and non-security-relevant commits. Then, we
created machine learning models based on the embeddings to predict the security
relevance of commits. Our results indicate no significant difference in the
embeddings between security-relevant and non-security-relevant commits when
looking at the features individually. However, we show that successful machine learning
models based on base classifiers and ensemble techniques can be trained on the
combination of the features. Compared to commit2vec, our model has a slightly
lower predictive performance, but it can predict a broader range of commits.
Finally, a model integrating both approaches achieves better results than those
obtained by the approaches independently.

As future work, we aim to extend our empirical study on a different dataset with
industrial and academic settings. We also plan to
experiment with other code analyzers and different ways of
processing their output. Since data about security-relevant commits is scarce,
we believe that transfer learning can be a valuable direction to explore, similarly
to the work of \citeauthor{lozoya2019commit2vec}~\cite{lozoya2019commit2vec}.
Finally, embeddings based on static code analyzers could be used for other
prediction tasks on source code such as automatic code repair and defect prediction.